\documentclass{styles/svproc}
\usepackage{url}
\usepackage{graphicx}

\begin{document}
\mainmatter              
\title{ChemDCAT-AP: Enabling Semantic Interoperability with a Contextual Extension of DCAT-AP}
\titlerunning{An Extension of the DCAT Application Profile}  
%

\author{Philip Strömert\inst{1} \and
Hendrik Borgelt\inst{2} \and
David Linke\inst{3} \and Mark Doerr\inst{4} \and Bhavin Katabathuni\inst{1} \and Oliver Koepler\inst{1} \and Norbert Kockmann\inst{2}}
\authorrunning{P. Strömert et al.} 
%
\tocauthor{Philip Strömert, Hendrik Borgelt, David Linke, Mark Doerr, Bhavin Katabathuni, Oliver Koepler, and Norbert Kockmann}
%
\institute{TIB - Leibniz Information Centre for Science and Technology, Germany 
\email{philip.stroemert@tib.eu}\and
TU Dortmund University, Germany\\
\and
 Leibniz Institute for Catalysis\\
 \and
 University of Greifswald\\}

\maketitle              

\begin{abstract}
Cross-domain data integration drives interdisciplinary data reuse and knowledge transfer across domains. However, each discipline maintains its own metadata schemas and domain ontologies, employing distinct conceptual models and application profiles, which complicates semantic interoperability. The W3C Data Catalog Vocabulary (DCAT) offers a widely adopted RDF vocabulary for describing datasets and their distributions, but its core model is intentionally lightweight. Numerous domain-specific application profiles have emerged to enrich DCAT’s expressivity, the most well-known DCAT-AP for public data. To facilitate cross-domain interoperability for research data, we propose DCAT-AP Plus, an Application \textbf{P}rofile \textbf{L}inking to \textbf{U}se-case \textbf{S}pecific context (\textbf{DCAT-AP+}). This generic application profile enables a comprehensive representation of the provenance and context of research data generation. DACT-AP+ introduces an upper-level layer that can be specialized by individual domains without sacrificing compatibility. We demonstrate the application of DCAT-AP+ and a specific profile \linebreak\textbf{ChemDCAT-AP} to showcase the potential of data integration of the neighboring disciplines chemistry and catalysis. We adopt LinkML, a YAML-based modeling framework, to support schema inheritance, generate domain-specific subschemas, and provide mechanisms for data type harmonization, validation, and format conversion, ensuring smooth integration of DCAT-AP+ and chemDCAT-AP within existing data infrastructures.

\end{abstract}

\section{Introduction}
Over the past decade, the research landscape has experienced a profound shift towards sharing of open data on an unprecedented scale. Initiatives like the European Open Science Cloud (EOSC) \cite{eosc2019strategicplan} and Germany’s National Research Data Infrastructure (NFDI) \cite{Hartl2021-cs} foster reproducible, efficient, and interdisciplinary research by making publicly funded research data.  With the growing adoption of the FAIR principles (Findable, Accessible, Interoperable, Reusable) \cite{ref_FAIR}, researchers now expect research data to be discoverable, reusable by minimum information standards that captures i.e. in chemistry or catalysis the full experimental context, including instrumentation, reaction conditions, provenance, and analytical methods, so results can be understood, reproduced, and combined across disciplines \cite{Herres-Pawlis2022-ol}. For the scientific communities in chemistry and catalysis the Physical Sciences Data Infrastructure (PSDI) \cite{PSDI_2023} initiative, the NFDI4Chem \cite{Steinbeck2020-oh} and NFDI4Cat \cite{NFDI4CAT-horsch} consortia address the development of minimum information standards.

The Data Catalogue Application Profile (DCAT-AP) is a successful example providing a robust foundation for metadata enabling data interoperability across European data portals \cite{DCAT-AP}. Nevertheless the detailed scientific context, what a dataset is truly about and how the data was generated are not covered by DCAT-AP specifications. For a chemistry or catalysis dataset such domain specific additions are needed for chemical structure notations (e.g., via InChI \cite{InChI2015} or SMILES \cite{SMILES}), reaction conditions, sample provenance, and specific analytical methods (e.g., NMR spectroscopy or gas chromatography).

To address this gap, NFDI4Chem and NFDI4Cat collaboratively developed the Chemistry DCAT Application Profile (\textbf{ChemDCAT-AP}).This profile serves as a common foundation for both consortia and is built upon a domain-agnostic core extension of DCAT-AP, called \textbf{DCAT-AP P}lus \textbf{L}inks to \textbf{U}se-case \textbf{S}pecific Context (\textbf{DCAT-AP+}). This core extension formalizes the already present but under-specified provenance pattern of DCAT-AP that is based on core classes from the Provenance Ontology (PROV-O)\cite{PROV_O}. It also adds a lightweight classification mechanism as well as a generic attribute pattern that can attach arbitrary quantitative or qualitative properties for the newly introduced classes. ChemDCAT‑AP imports DCAT‑AP+ and adds chemistry‑ and catalysis‑specific concepts (chemical entities, reactions, analytical methods) aligned with established chemistry ontologies (i.e., ChEBI, CHEMINF, and CHMO) \cite{EBI_CHEBI,CHEMINF,CHMO} thereby offering a shared semantically grounded metadata foundation for NFDI4Chem and NFDI4Cat.

Our  framework leverages the Linked Data Modeling Language (LinkML) \cite{ref_linkml}, a flexible modeling framework that allows us to create a modular schema where the foundational ChemDCAT-AP can be easily imported and extended to  specific needs of chemistry and catalysis. This approach makes semantic standards more accessible by compiling the schema into various practical formats, including Python/Pydantic data classes, SHACL shapes and a JSON
Schema, thereby bridging the gap between semantic rigor and practical implementation for repository developers.
The main contributions of this article are:
\begin{itemize}
    \item Translation of DCAT-AP 3.0 specification into a LinkML schema.
    \item Design of the domain-agnostic extension \textbf{DCAT-AP+} to allow a more provenance-based description of how a dataset was generated and what it is about.
    \item \textbf{ChemDCAT-AP} as a shared foundational profile that imports and specializes DCAT-AP+ for the chemistry and catalysis domains.
    \item Prototypical implementation of a Minimum Information for a Chemical Investigation (MIChI) standard for \textbf{NMR spectroscopy} demonstrating \linebreak ChemDCAT-AP's adaptability experimental contexts as well as its implementation within the NFDI4Chem Search Service.
\end{itemize}

This paper is organized as follows: Section 2 presents the background on relevant metadata standards and the LinkML framework. Section 3 outlines the methodology used to design and build our extension. Section 4 presents the results, detailing the architecture of ChemDCAT-AP and its implementation in the NFDI4Chem Search Service. Finally, Section 5 discusses the implications and limitations of our work, followed by conclusions and future directions.

\section{Background}
\subsection{The Need for Domain-Specific Metadata}

With the growth of Linked Open Data, the need for domain-specific metadata has become increasingly clear. Standards such as DataCite \cite{DataCiteMetadataSchema4.6}, Dublin Core \cite{DublinCore}, and METS \cite{METS} support dataset description in XML, JSON, or RDF, while W3C vocabularies such as DCAT \cite{ref_dcat_voc}, PROV-O \cite{PROV_O}, and the Semantic Sensor Network Ontology (SOSA \cite{ref_sosa} add a semantic layer. Together, these enable machine-readable metadata across domains.

Yet such general frameworks remain too generic to capture detailed scientific context. The CDIF (Cross Domain Interoperability Framework) \cite{ref_cdif} highlights the gap between standards supporting discoverability (e.g., schema.org \cite{schema_org}, DCAT) and those addressing content or variables (e.g., SOSA, I-ADOPT \cite{ref_i_adopt}). Bridging this gap is particularly important in chemistry, catalysis, and materials research, where datasets are deeply interlinked through reactions, catalysts, and experimental conditions.

\subsection{DCAT-AP: An Ontology-Aligned Metadata Schema}

DCAT-AP, the European application profile of DCAT \cite{W3C-DCAT3}, provides a common specification for describing public sector datasets across portals while allowing catalogs to maintain their own storage and documentation. It ensures semantic interoperability by aligning with controlled vocabularies, taxonomies, and ontologies.
\\
Extensions of DCAT-AP must follow strict implementation guidelines \cite{DCATAP_Guidelines}:
\begin{itemize}
\item Usage notes may only be narrowed, not broadened.
\item New classes and properties may be added, but not if they duplicate existing ones.
\item Cardinalities for properties can be modified within limits, but mandatory requirements must remain intact.
\end{itemize}

As an application profile, DCAT-AP constrains the general DCAT vocabulary without redefining concepts. Expressed in SHACL, it regulates the use of ontology terms while ensuring interoperability across different profiles \cite{doi:10.3233/SW-233156}.

\subsection{Existing DCAT-AP Extensions}

Several domain-specific profiles illustrate the flexibility of DCAT-AP:

\textbf{GeoDCAT-AP} \cite{ref_geodcat-ap-3.0.0} extends DCAT-AP to describe geospatial datasets, series, and services. It addresses metadata needs specific to geographic information, improving discovery and reuse across European data portals. Key features include richer metadata on agents involved in dataset creation, processing, and use, as well as a few domain-specific properties, such as \textit{spatial\_resolution\_as\_text} and \textit{topic\_category}. As an official extension, it follows the same methodology as DCAT-AP to ensure interoperability.

\textbf{StatDCAT-AP} \cite{statdcat} extends DCAT-AP for statistical datasets, enabling interoperability both within the statistical domain and with open data portals. Version 1.0.1 supports metadata for formats such as Statistical Data and Metadata eXchange specification (SDMX), RDF Data Cube, and CSV. A defining feature is its ability to describe multidimensional structures, capturing variables across geographic, temporal, and domain-specific dimensions. Developed through the same Working Group and review process as DCAT-AP, it ensures broad stakeholder engagement and compatibility.

\textbf{MLDCAT-AP} \cite{MLDCAT-AP_2.0.0} extends DCAT-AP for machine learning models, their datasets, quality measures, and related publications. It incorporates specialized vocabularies, including mls and mlso (machine learning), biro (citations), dqv (data quality), and lpwc/lpwc-p (papers with code). Still in Candidate Recommendation status, MLDCAT-AP represents early work toward standardizing metadata for AI assets, focusing on model–dataset relationships rather than dataset content.

\textbf{HealthDCAT-AP} \cite{healthDCAT-AP} extends DCAT-AP for health datasets, addressing domain-specific requirements and ensuring alignment with the European Health Data Space (EHDS) regulation. Its purpose is to improve interoperability and discoverability of health-related data across Europe.

These extensions primarily add properties to the Dataset class but do not share a common design pattern. The DCAT-AP+ approach proposed here aims to harmonize such efforts through a provenance-based model, aligning with suggestions by DCAT-AP developers \cite{look_through_DCATAP}.

\subsection{Alternatives to DCAT-AP}

For describing data content in a discoverability context, two prominent alternatives to DCAT and DCAT-AP are Schema.org and DataCite.

\textbf{Schema.org}, developed by major search engine providers, embeds metadata directly into websites to support search functionalities. While it can describe content similar to DCAT, it is optimized for web indexing, offers limited expressivity for research data, and is less flexible to extend. A notable domain-specific extension is Bioschemas for life sciences data.

\textbf{DataCite} \cite{DataCiteMetadataSchema4.6} provides metadata linked to Digital Object Identifiers (DOIs), ensuring stable, persistent citation and discovery of research outputs. However, its metadata is tightly bound to this persistent identifier (PID) infrastructure, making adaptations more difficult than with open models like DCAT-AP.


Both serve important roles but are not sufficient for detailed, domain-specific metadata.

\subsection{The LinkML Framework: A Foundational Choice}
\label{The LinkML Framework: A Foundational Choice}

The Linked Data Modeling Language (LinkML) was chosen as the implementation framework because it simplifies the creation of FAIR-compliant standards while making Semantic Web technologies more approachable. Its YAML-based syntax enables schema design in a single source of truth, from which multiple machine-interpretable representations can be generated (Python/Pydantic classes, JSON Schema with JSON-LD, SHACL shapes, Markdown).

Schema elements (classes, properties, enumerations, and datatypes) can be mapped to external artifacts such as OWL ontologies or SKOS thesauri via their term IRIs. This anchors a LinkML schema in shared semantics and promotes interoperability. Because these mappings are optional, users can benefit from the Semantic Web tool stack without being locked into it; other auto-generated representations may already provide sufficient standardization for many use cases.  Yet when this feature is used, LinkML's data conversion libraries can become a very accessible tool to quickly generate RDF knowledge graphs. 

By swapping once chosen term IRI mappings for equivalent ones, a single LinkML schema can integrate RDF data built on different ontological commitments, as long as the used ontologies follow the same modeling patterns. For example, a \verb|Person| class with attributes \verb|given_name| and \verb|family_name| can be aligned with either the FOAF ontology \cite{BrickleyMiller2010_FOAF} or Schema.org, since both define equivalent structures.


This versatility, combined with its practical tooling and active community, makes LinkML well suited for extending DCAT-AP and for adoption by diverse user groups, from developers to data stewards.

\section{Methods}

This section outlines the design rationale, technical implementation, and validation process employed in the development of \textbf{ChemDCAT-AP}. Its collaborative and iterative development process followed several core principles:

\begin{enumerate}
    \item \textbf{Conformance}: All extensions strictly adhere to the mandatory constraints of the official DCAT-AP 3.0 specification.
    \item \textbf{Discoverability-centric}: New properties are chosen to directly improve dataset findability for specific chemistry and catalysis use cases.
    \item \textbf{Semantic grounding}: Every added class or property is mapped to an established ontology (e.g., PROV-O, QUDT, BFO, IAO, OBI, CHMO) \cite{PROV_overview,QUDT,ref_BFO,IAO,OBI}.
    \item \textbf{Simplicity for non-specialists}: The schema remains usable by data stewards and software developers without requiring deep ontology expertise, while still providing a rich, machine-readable representation.
\end{enumerate}

The concrete workflow that transformed these principles into the final \linebreak \textbf{LinkML} model followed three tightly coupled stages: (i) the faithful automatic porting of the DCAT-AP into LinkML, (ii) the systematic extension of that DCAT-AP LinkML schema with a provenance pattern, and (iii) the continuous integration-driven development of the ChemDCAT-AP schema. The source code, documentation, and all model artifacts are publicly hosted on GitHub\footnote{https://github.com/nfdi-de/chem-dcat-ap}.

\subsection{Modeling Framework and Implementation}

The main reasons for choosing LinkML as our implementing framework were laid out in detail in \ref{The LinkML Framework: A Foundational Choice}. Bootstrapping the project with the official LinkML project copier template\footnote{https://github.com/linkml/linkml-project-copier} provided us a robust starting point with many benefits, such as:
\begin{itemize}
    \item a standardized repository layout including an example schema and test file,
    \item pre-configured GitHub Actions for continuous integration (CI) and quality control,
    \item or a  \verb|just| command runner containing standardized commands for validating schemas and data examples, testing and deploying the HTML documentation\footnote{https://nfdi-de.github.io/chem-dcat-ap/}, or updating the virtual Python environment and template code
\end{itemize}
The complete tool-chain relies on Python 3, leveraging libraries such as \linebreak \verb|linkml-runtime| for data handling, \verb|pydantic| for data validation, and \verb|rdflib| for RDF manipulation.

\subsection{Porting DCAT-AP to LinkML}

To ensure conformance with our first guiding principle, we developed a Python script to port the DCAT-AP 3.0.0 SHACL shapes to LinkML. We used the \textbf{JSON-LD} serialization of the DCAT-AP specification as our ground truth, as it allowed for simpler parsing with Python's built-in JSON library. We specifically used the version from the \verb|master| branch of the DCAT-AP GitHub repository, which is more recent than the release tagged with version 3.0.0 and because it is the version linked from the official specification website.\footnote{see also https://github.com/SEMICeu/DCAT-AP/issues/428}

The script reads the JSON-LD file and translates each SHACL node shape into a LinkML construct. Node shapes with a \verb|targetClass| pointing to an ontology class become LinkML \textbf{classes}, while those designating an XSD datatype become LinkML \textbf{datatypes}. Property shapes within these nodes are interpreted as the slots of the derived LinkML classes. The original term IRIs from the SHACL \verb|targetClass| and \verb|path| attributes are preserved verbatim as the \verb|class_uri| and \verb|slot_uri|, ensuring the resulting model is semantically identical to the original SHACL representation.

A particular challenge arose from the presence of \textbf{union ranges} in the original SHACL shapes. For properties like \verb|dcat:primaryTopic|, LinkML's \verb|any_of| keyword was used to support unions of object classes. However, due to the lack of stable support for datatype unions in the current LinkML implementation, we conservatively restricted the range of date-related slots to the XSD datatype \verb|date|. This is a deliberate, stricter interpretation that is acceptable for our current use cases while we await a future LinkML release with full datatype union support.

\subsection{A Two-Layered Modular Design}

We adopted a modular, two-layered design that first leveraged the same Python script to programmatically extend our faithful LinkML translation of DCAT-AP into DCAT-AP+, using the \verb|wasGeneratedBy| property and its range class \verb|Activity| as point of entry. In a second step, DCAT-AP+ was imported into the manually created ChemDCAT-AP. 

\subsubsection{The Provenance Core Layer}

The \textbf{DCAT-AP+} layer is a minimal, domain-agnostic extension that formalizes the description of data provenance. It introduces a generic pattern for describing data generation activities, aligning with the W3C PROV Ontology (PROV-O) and other established vocabularies. This choice was further validated by Prudhomme et. al. \cite{Prudhomme2025} , who described a formal mapping between PROV and the Basic Formal Ontology (BFO), which aligns with our use of OBO Foundry ontologies within NFDI4Chem and NFDI4Cat.

We first implemented the core PROV-O activity pattern by extending the in DCAT-AP under-specified \verb|Activity| class with its PROV-O properties (renamed for clarity to: \verb|has_input_entity|, \verb|has_output_entity|, \verb|carried_out_by| and \verb|has_input_activity|) and their respective not yet in DCAT-AP present range classes \verb|Entity| and \verb|AgentEntity|. Of which the latter was renamed due to the \verb|foaf:Agent| class being already used for a different purpose in DCAT-AP. 
We then specialized this core pattern with a \verb|DataGeneratingActivity| class, a subclass of \verb|Activity|, to represent processes like a measurement or simulation. This activity has as main input an \verb|EvaluatedEntity| (e.g., a chemical sample) or a \verb|EvaluatedActivtiy| (e.g., a catalytic reaction), which is described using the dedicated sub-properties \verb|evaluated_entity| and \verb|evaluated_activity|. The \verb|carried_out_by| such an activity (e.g., a reactor) can be further specialized using one of the \verb|AgentEntity| subclasses \verb|Software| or \verb|Device|.

To describe entities, activities or agents in detail, we added classes for qualitative and quantitative attributes. The \verb|QuantitativeAttribute| class (aligned with \verb|qudt:Quantity|) includes properties for numerical values, quantity kind, and units. The \verb|QualititativeAttribute| class (aligned with \verb|prov:Entity|) allows to provide a string value for its literal representation. They are used via the \verb|has_quantitative_attribute| and \verb|has_qualitative_attribute| properties, which both are mapped to \verb|dcterms:relation|.

A crucial feature of DCAT-AP+ is the \verb|ClassifierMixin|, which provides two properties for semantic annotation to all DCAT-AP+ core classes:
    \begin{itemize}
        \item \verb|rdf_type|: Mapped to the RDF property \verb|rdf:type|, this allows an instance to be additionally typed with a class from an external ontology (e.g., classifying a \verb|DataGeneratingActivity| instance with \textit{CHMO:0000595} for a ¹³C NMR measurement). An approach endorsed by the DCAT-AP developers \cite{look_through_DCATAP}
        \item \verb|type|: Mapped to \verb|dcterms:type|, this allows for classification using a term from SKOS vocabularies or other controlled term sets without making such a strong ontological commitment as with \verb|rdf_type|.
    \end{itemize}

\begin{figure}
\includegraphics[width=1\textwidth]{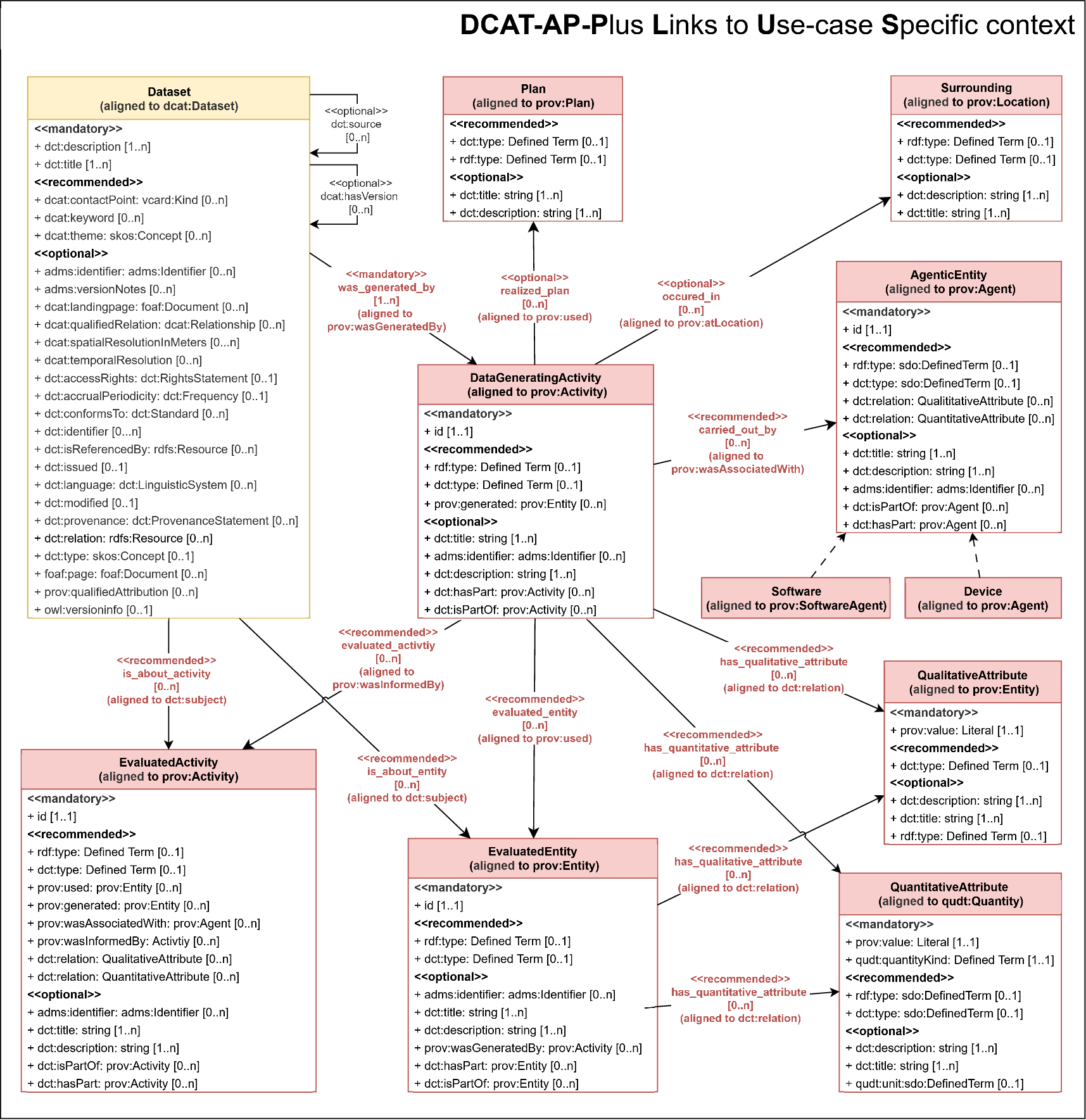}
\caption{This visualization depicts how we extended the DCAT-AP via the \textit{prov:Activity} class, to be able to provide additional domain-specific metadata for the entities or activities that are the main subject of a \textit{dcat:Dataset}, as well as for the instruments and plan that were used within the creation of a dataset and the surrounding in which this activity took place. Also depicted are the mappings we used to ground this LinkML schema within the ontological commitment of  PROV-O, DCTerms and QUDT.\cite{QUDT}} \label{fig1}
\end{figure}

For optional metadata, the DCAT-AP+ reuses standard DCAT-AP and Dublin Core properties to allow providing identifiers, labels, descriptions, and part-whole relations. It also reuses the PROV-O classes \verb|Plan| and \verb|Location| in a sparsely specified way, which can be seen in more detail in the DCAT-AP+ UML diagram depicted by Figure \ref{fig1}.

\subsubsection{The Domain-Specific Layer}

The \textbf{ChemDCAT-AP} layer serves as the \linebreak domain-specific extension that tailors the generic provenance model of DCAT-AP+ to the explicit needs of the chemistry and catalysis research domains. It achieves this by importing all entities from DCAT-AP+ and introducing specialized classes and properties that are directly mapped to established chemical ontologies. This layered approach ensures that metadata remains fully compliant with DCAT-AP+ and thereby also with the base DCAT-AP standard while gaining the semantic precision necessary to support advanced, domain-specific discovery use cases. The primary design goal of ChemDCAT-AP is to move from generic key-value descriptions to a rich, self-describing data model that reflects the core entities of chemical research, such as substances, samples, and reactions.

A key strategy in ChemDCAT-AP is the specialization of generic classes from DCAT-AP+. The generic \verb|EvaluatedEntity| is subclassed to create chemically meaningful concepts like \verb|SubstanceSample| (mapped to \verb|SIO:001378|) and \verb|ChemicalEntity| (mapped to \verb|CHEBI:23367|). This allows for the direct representation of a chemical sample or a molecular entity within the metadata record, removing the ambiguity of the generic model. For instance, a \verb|SubstanceSample| is further refined with the \verb|ChemicalSubstance| mixin, which provides properties to describe its composition (\verb|composed_of|) and key chemical attributes like concentration (\verb|has_concentration|) or pH (\verb|has_ph_value|).

Furthermore, ChemDCAT-AP introduces a rich set of domain-specific properties that are defined as sub-properties of their generic DCAT-AP+ counterparts. This aligns with the recommendations for creating application profiles, ensuring a clear inheritance model \cite{DCATAP_Guidelines,look_through_DCATAP}. For example, to describe the components of a chemical reaction, ChemDCAT-AP provides highly specific properties. The property \verb|used_catalyst| is defined as a sub-property of the generic \verb|carried_out_by| but is more precisely mapped to the predicate \verb|RXNO:0000425| ('has\_catalyst') from the Named Reaction Ontology (RXNO) \cite{RXNO-Ontology}. Similarly, the property \verb|generated_product| specializes \verb|had_output_entity| and is mapped to \verb|RO:0004008|. This hierarchical property design provides richer, domain-specific semantics while maintaining full backward compatibility with the more generic DCAT-AP+ model. Any client capable of understanding DCAT-AP+ can still interpret \verb|used_catalyst| as a form of agent involved in the activity, while a chemistry-aware client can understand its precise role as a catalyst.

The practical benefit of this specialization is evident when comparing a metadata record for a chemical substance sample under both models.\footnote{https://github.com/nfdi-de/chem-dcat-ap/blob/main/tests/data/valid/NMRAnalysisDataset-001.yaml} As shown in the provided examples, a record conforming to DCAT-AP+ must rely on the generic \verb|has_qualitative_attribute| property and use the \verb|rdf_type| slot to specify that a particular string represents an InChIKey (\verb|CHEMINF:000059|) or a SMILES string (\verb|CHEMINF:000018|). In contrast, the ChemDCAT-AP-conformant record utilizes the dedicated \verb|inchikey| and \verb|smiles| properties. These properties have the CHEMINF ontology classes defined as their range within the LinkML schema itself, thus eliminating the need for repeated \verb|rdf_type| annotations in the instance data. This not only makes the instance data more concise and readable but also greatly simplifies the process of mapping data from existing chemistry software and databases, which often have similarly named fields. By providing a more explicit and constrained model, ChemDCAT-AP reduces the potential for ambiguity and enhances the potential for automated data validation and integration into a larger Chemistry Knowledge Graph.

\subsection{Validation and Application}

The syntactic correctness of the ChemDCAT-AP schema was continuously ensured through an automated testing pipeline executed via GitHub Actions on every pull request. This pipeline utilized the \verb|linkml-validate| tool to check the integrity of the LinkML schema itself and to validate example data instances against it.

To assess its fitness for purpose, we applied the model to represent real-world metadata from the \textbf{Chemotion Repository}, an electronic lab notebook and repository widely used within the NFDI4Chem consortium. We successfully mapped all existing metadata fields from Chemotion records to the \linebreak ChemDCAT-AP schema. This evaluation confirmed that ChemDCAT-AP not only accommodates all current metadata but also provides the necessary structure to capture more detailed semantic information, thereby meeting our primary design goals.

\section{Results}
Our primary result is a flexible, two-layered metadata framework that successfully bridges the gap between generic data cataloging and the specific descriptive needs of chemical research. We demonstrate its efficacy through two key accomplishments: the formalization of a Minimum Information for a Chemical \linebreak Investigation (MIChI) standard for NMR spectroscopy, and the framework's successful implementation within the NFDI4Chem Search Service.

\subsection{Formalizing the NMR MIChI with ChemDCAT-AP}
A key achievement of our work is the successful formalization of the NFDI4Chem MIChIs (Minimum Information about Chemical Investigation) for NMR spectroscopy, known as MARGARITAS \cite{margaritas}. The ChemDCAT-AP framework proved highly effective, allowing for a collaborative and iterative development process where the main challenges were the correct interpretation of the standard's semantics, not the technical modeling itself.

This process resulted in the formalization of all required and most recommended properties within the MARGARITAS standard. Notably, our model was able to capture richer metadata than is currently exposed via the API of the Chemotion repository. Details, such as the calibration compound or the probed nucleus, while present in raw data files or landing pages, could be structured for the first time using our model.

Beyond the schema itself, a significant result is also the suite of auto-generated artifacts produced by the LinkML toolchain. The human-readable documentation facilitates review and contribution from domain experts, while the multiple machine-readable representations (e.g., Python/Pydantic, JSON Schema) lower the barrier to adoption for repository managers and software developers seeking to align with this formalized version of the MARGARITAS. The entire workflow now serves as a validated template for systematically encoding other MIChI standards, promising to accelerate the creation of FAIR chemical research data.

\subsection{Implementation in the NFDI4Chem Search Service}

The practical utility of ChemDCAT-AP is confirmed by its successful integration into the NFDI4Chem Search Service\footnote{https://search.nfdi4chem.de/}, which is built on CKAN \cite{ref_ckan}. By developing a custom profile for the existing \textbf{ckanext-dcat} plugin\footnote{https://github.com/ckan/ckanext-dcat}, we enabled the service to parse and export rich, ChemDCAT-AP-compliant metadata. The auto-generated Python data classes from our LinkML model were instrumental in this process, providing a straightforward mechanism for validating and transforming metadata harvested by the Search Service into a standardized RDF graph.

This implementation is not merely a technical exercise; it directly enables the creation of the NFDI Chemistry Knowledge Graph (ChemKG). The semantically precise metadata captured by our model allows for advanced, faceted search queries that are impossible with generic catalog metadata, such as finding all datasets where a specific solvent was used or that analyze a compound with a given molecular scaffold. The ability to export this structured metadata directly from the Search Service and have it served via its SPARQL endpoint is a significant step toward greater interoperability across research infrastructures.

\section{Discussion}
We have demonstrated that the layered extension of DCAT-AP, realized through DCAT-AP+ and ChemDCAT-AP, offers a robust and scalable solution for describing domain-specific research data. The provenance-based pattern at the core of DCAT-AP+ provides a harmonized design principle that other domains could adopt, while ChemDCAT-AP confirms its suitability for the complex requirements of chemistry and catalysis.

While this work establishes a solid foundation, several technical and strategic avenues remain for future development. On the technical side, our immediate priority is to publish the schemas under persistent URLs (PURLs) via w3id.org and release the models as a package on PyPI to encourage broader reuse. We also plan to separate DCAT-AP+ into a standalone repository to facilitate its adoption by other NFDI consortia and interested parties. Furthermore, our work has identified limitations in LinkML's current SHACL generator concerning the handling of IRI naming for node and property shapes. We intend to contribute to the LinkML community to address this, ensuring that the generated shapes are fully compliant with the current SHACL specifications.

Strategically, the long-term vision is for DCAT-AP+ to be considered for integration into a future release of the official DCAT-AP standard, as it provides a logical and needed formalization of the \verb|prov:wasGeneratedBy| property. However, this requires thorough validation and adoption by a wider community. We have initiated this process by gathering feedback within the NFDI network and extend an invitation to the broader research data community to test and critique DCAT-AP+.

Ultimately, the ChemDCAT-AP framework serves as a cornerstone for the planned Metadata Schema Service (MSS) within the second funding phase of NFDI4Chem. This service will not only host chemistry-specific and other relevant schemas but also provide mappings and automated transformations between them, a task for which the multiple code representations of ChemDCAT-AP are ideally suited. By integrating it with the currently developed EOSC Metadata Schema and Crosswalk Registry (MSCR) \cite{FAIRCORE4EOSC_MSCR}, this work contributes to a more interconnected and semantically interoperable research data ecosystem.

\subsubsection{\ackname}
This research was funded by the German Research Foundation (DFG), grant number 441926934 (NFDI4Cat)  and grant number 441958208 \linebreak (NFDI4Chem).

\end{document}